\documentclass[12pt]{article}

\newcommand{\be}{\begin{equation}}
\newcommand{\ee}{\end{equation}}
\newcommand{\beqs}{\begin{eqnarray}}
\newcommand{\eeqs}{\end{eqnarray}}
\def\pbfp{{\hbox{\scriptsize+}
 \kern-1.32ex\hbox{\raise0.05ex\hbox{\scriptsize+}}
 \kern-1.35ex\hbox{\raise0.03ex\hbox{\scriptsize+}}
 \kern-1.24ex\hbox{\lower0.03ex\hbox{\scriptsize+}}
 \kern-1.26ex\hbox{\lower0.05ex\hbox{\scriptsize+}}}}
\def\pbfph{{\raise0.36ex\hbox{\scriptsize+}
 \kern-1.32ex\hbox{\raise0.41ex\hbox{\scriptsize+}}
 \kern-1.35ex\hbox{\raise0.39ex\hbox{\scriptsize+}}
 \kern-1.24ex\hbox{\raise0.33ex\hbox{\scriptsize+}}
 \kern-1.26ex\hbox{\raise0.31ex\hbox{\scriptsize+}}}}
\def\pbfpf{{\hbox{\scriptsize+}
 \kern-1.57ex\hbox{\raise0.03ex\hbox{\scriptsize+}}
 \kern-1.55ex\hbox{\raise0.05ex\hbox{\scriptsize+}}
 \kern-1.51ex\hbox{\raise0.05ex\hbox{\scriptsize+}}
 \kern-1.49ex\hbox{\lower0.03ex\hbox{\scriptsize+}}}}
\def\pbfm{{\hbox{\Large-}
 \kern-1.01ex\hbox{\raise0.06ex\hbox{\Large-}}}}
\def\pbfmf{{\hbox{\large-}
 \kern-1.1ex\hbox{\raise0.04ex\hbox{\large-}}}}
\def\pbfmm{{\lower.36ex\hbox{\Large-}
 \kern-1.01ex\hbox{\lower0.3ex\hbox{\Large-}}}}
\def\bpm{{\hbox{\scriptsize+}
 \kern-1.32ex\hbox{\raise0.05ex\hbox{\scriptsize+}}
 \kern-1.35ex\hbox{\raise0.03ex\hbox{\scriptsize+}}
 \kern-1.24ex\hbox{\lower0.03ex\hbox{\scriptsize+}}
 \kern-1.26ex\hbox{\lower0.05ex\hbox{\scriptsize+}}
 \kern-1.01ex\hbox{\lower1.0ex\hbox{\Large-}}
 \kern-1.01ex\hbox{\lower1.06ex\hbox{\Large-}}}}
\def\({\left(}
\def\){\right)}

\def\th{\theta}

\def\pap{{\pa_{\pbfp}}}
\def\papf{{\pa_{\pbfpf}}}
\def\pam{{\pa_{\pbfm}}}

\def\cp{{c^{\pbfp}}}

\def\cmin{{c^{\pbfmm}}}

\def\bp{{b_{\pbfp\pbfp}}}

\def\bm{{b_{\pbfm\pbfm}}}

\def\bep{{\beta_{\pbfp+}}}
\def\bem{{\beta_{\pbfmm-}}}
\def\Lag{ {\cal L}}

\def\Dl{ {\cal D}}

\def\mxth{\mathsurround=0pt }
\def\xversim#1#2{\lower2.pt\vbox{\baselineskip0pt \lineskip-.5pt
x  \ialign{$\mxth#1\hfil##\hfil$\crcr#2\crcr\sim\crcr}}}

\newcommand{\half}{{\textstyle{\frac12}}}
\renewcommand{\a}{\alpha}
\renewcommand{\b}{\beta}
\renewcommand{\c}{\gamma}
\renewcommand{\d}{\delta}
\newcommand{\D}{\Delta}
\newcommand{\pa}{\partial}
\newcommand{\g}{\gamma}
\newcommand{\G}{\Gamma}
\newcommand{\e}{\epsilon}

\renewcommand{\k}{\kappa}
\renewcommand{\l}{\lambda}
\renewcommand{\L}{\Lambda}
\newcommand{\LB}{{\cal K}}

\newcommand{\pis}{{\pi\kern-1.28ex /}}

\newcommand{\s}{\sigma}
\renewcommand{\S}{\Sigma}

\renewcommand{\o}{\omega}
\newcommand{\q}{\theta}
\newcommand{\h}{\eta}

\newcommand{\ft}[2]{{\textstyle\frac{#1}{#2}}}

\newcommand{\eqn}[1]{(\ref{#1})}

\begin{document}
\begin{titlepage}
\renewcommand{\thefootnote}{\fnsymbol{footnote}}
\begin{center}
\hfill ITP-UU-01/12  \\
\hfill YITP-01/18  \\
\hfill {\tt hep-th/0211266}\\
\vskip 20mm

{\Large {\bf Consistent Boundary Conditions for Open Strings}}

\vskip 10mm
{\bf Ulf Lindstr\" om\footnote{Permanent address: {\em Department of Theoretical Physics, Uppsala University, P.O.Box 803, S-75108, Uppsala Sweden.}},
Martin Ro\v{c}ek, and Peter van Nieuwenhuizen }
\vskip 4mm

{\em C.N. Yang Institute for Theoretical Physics}\\
{\em SUNY, Stony Brook, NY 11794-3840, USA}\\ 
{\tt ul@physto.se}\\
{\tt rocek@insti.physics.sunysb.edu}\\
{\tt vannieu@insti.physics.sunysb.edu}
\vskip 6mm

\end{center}
\vskip .2in

\begin{center} {\bf ABSTRACT } \end{center}
\begin{quotation}\noindent
We study boundary conditions for the bosonic, spinning (NSR) and Green-Schwarz open string, as well as for $1+1$ dimensional supergravity. We consider boundary conditions that arise from (1) extremizing the action, (2) BRST, rigid or local supersymmetry, or $\k$(Siegel)-symmetry of the action, (3) closure of the set of boundary conditions under the symmetry transformations, and  (4) the boundary limits of bulk Euler-Lagrange equations that are ``conjugate'' to other boundary conditions.  We find corrections to Neumann boundary conditions in the presence of a bulk tachyon field. We discuss a boundary superspace formalism. We also find that path integral quantization of the open string requires an infinite tower of boundary conditions that can be interpreted as a smoothness condition on the doubled interval; we interpret this to mean that for a path-integral formulation of open strings with only Neuman boundary conditions, the description in terms of orientifolds is not just natural, but is actually fundamental. 

\end{quotation}
\vfill
\flushleft{\today}
\end{titlepage}
\renewcommand{\thefootnote}{\arabic{footnote}}
\setcounter{footnote}{0}
\eject
\section{Introduction and Summary}
\setcounter{equation}{0}
We  study boundary conditions in open string theory, or more
generally, in field theories on $1+1$ dimensional spacetimes with boundaries. Four kinds of boundary conditions arise: (1) Conditions needed to extremize the action. (2) Conditions needed to ensure invariance of the action under some symmetry transformations. (3) Conditions needed to ensure that the boundary conditions themselves are closed under the action of the symmetry.  (4) Conditions that  are ``conjugate'' to other boundary conditions in the following sense: When we vary the action to find the equations of motion, we have variations of the form $\d\phi^i\frac\pa{\pa\phi^i}\Lag$, where $\phi^i$ represents some generic set of fields; if we consider the limit of this as we approach the boundary, then the boundary conditions may restrict some combination of $\d\phi^i$; thus, some corresponding conjugate combination of the field equations $\frac\pa{\pa\phi^i}\Lag$ is not implied by the variational principle on the boundary, and must be imposed by hand. 

Conditions (1) and (2) have been used since the early literature on BRST quantization of the open string \cite{pvn}, but conditions of type (3) frequently have been overlooked in the literature, and were only recently discussed in \cite{hlz,alz,alz2,lz} (see, however, \cite{hiv}). They are a logical necessity: A theory with fields $\phi$ is invariant under a symmetry $\phi\to\tilde\phi$ if nothing changes when it is written in terms of the transformed fields $\tilde\phi$. This in particular implies that one cannot distinguish whether the boundary conditions are imposed on $\phi$ or $\tilde\phi$, which implies condition (3). Conditions (4) are conceptually entirely new\footnote{In the examples we consider here, they are often, but {\em not} always (see, {\it e.g.}, the case of Dirichlet boundary conditions in section 2), redundant with the conditions that arise from (1--3).}.

In general, imposing all four kinds of conditions leads to more boundary conditions than the usual two for a second-order field equation and one for a first-order field equation. These ``extra'' boundary conditions are not over-restrictive because they arise as the {\em restriction of some of the field equations to the boundary}. This guarantees that the field equations continue to have solutions even when there seem to be too many boundary conditions to specify the boundary-value  problem. Actually, we find that only those field equations that do not involve time derivatives of the boundary fields arise. Thus the Cauchy problem of the boundary fields is also unchanged. 

We consider several examples, including the BRST symmetry of the bosonic and spinning strings, the rigid and local worldsheet supersymmetry of the spinning string, and the $\k$(Siegel)-symmetry of the superstring. In particular, we find that Neumann boundary conditions receive nontrivial boundary corrections from a superpotential ({\it i.e.}, a bulk tachyon).  Such a term has appeared previously in the literature in the context of $N=2$ boundary integrable models \cite{w}. Strikingly, the boundary condition we find is compatible with the BPS equation.  We also give a superspace derivation of the boundary term. Some earlier work on open strings propagating in the presence of a tachyon background can be found in \cite{other1,other2,other3,other4}; none, however, consider the effects that we describe here.

The requirement that the boundary conditions of the open bosonic or spinning string close under BRST transformations leads to a surprising result.  When we impose BRST invariance on the boundary conditions, we are led to an infinite tower of boundary conditions: for instance, in addition to $\cp-\cmin=0$ and $\pa_\s(\cp+\cmin)=0$, we find $\pa_\s^2(\cp-\cmin)=0$, $\pa_\s^3(\cp+\cmin)=0$, etc., and similarly for the antighosts. This tower of conditions can be interpreted as requiring smoothness of a field, {\it e.g.}, $c(\s)$ defined on the double interval by $c(\s)=\cp(\s)$ for $\s>0$ and $c(\s)=\cmin(-\s)$ for $\s<0$. The analogous conditions on the coordinate $X^\mu(t,\s)$ imply that it is a symmetric function on the doubled interval. This has a modern interpretation: one may regard open strings as the ``twisted sector'' in the presence of a charge-neutral stack of a space-filling orientifold plane and 16 D-branes \cite{sag1,sag2,sag3}. Thus our results imply that for a path-integral formulation of open strings with only Neuman boundary conditions, the description in terms of orientifolds is not just natural but is actually fundamental. 

In some recent articles on cosmology, tachyons of the bosonic string have been proposed as a mechanism that produces the observed positive acceleration of the universe \cite{g}. In these models, no boundary terms of the kind discussed in this article were considered. It might be interesting to see if such terms lead to any qualitative changes.

Boundary conditions that preserve $\k$(Siegel)-symmetry  have
been discussed in the context of superembeddings in
\cite{chs}, and conditions that preserve the BRST
symmetry of the Berkovits string has been discussed in
\cite{bp}. In both these cases the boundary
conditions lead to Born-Infeld dynamics for the boundary spin
one field.

Recently, a covariant quantum superstring with a simpler BRST symmetry than would follow from the quantization of the classical $\k$(Siegel)-gauge symmetry has been constructed, and the boundary conditions that follow from the Euler-Lagrange equations, the BRST symmetry, and the rigid spacetime supersymmetry were derived \cite{gpvn}; it would be interesting to apply the full program presented in this paper to that case.

\section{Supersymmetry and superpotentials}
\setcounter{equation}{0}
We now study $(1+1)$-dimensional supersymmetric Minkowski-space field theories with a superpotential; such theories describe the propagation of NSR strings in a bulk tachyon background. These theories have also been studied in the context of supersymmetric solitons \cite{svv,nsrvn,glvn,bgvnv}.

Consider the supersymmetric action
\be
I=\frac1{\pi}\int_\S \left( \Lag_{\S}
+\Lag_{\pa\S}\right)~,
\label{act}
\ee
where
\be
\Lag_{\S}=2\pap X\pam X + i(\psi^+\pap \psi^+ + \psi^-\pam\psi^-) +
\frac12F^2+ FW'(X) +iW''(X)\psi^+\psi^-~.
\label{lag}
\ee
Here $W$ is an {\it arbitrary} superpotential, 
and we determine the boundary Lagrangian $\Lag_{\pa\S}\equiv\pa_\s\LB$ in terms of $W$. We may also write the boundary contribution to the action as
\be
\frac1{\pi}\int_\S \Lag_{\pa\S} = \frac1\pi\int_{\pa\S}\LB~.
\ee
In our conventions, $\pa_\s\equiv\pap-\pam$,  $\pa_t\equiv\pap+\pam$, so
$\pap=\frac12(\pa_t+\pa_\s)$ and $\pam=\frac12(\pa_t-\pa_\s)$.

Varying with respect to the various fields in the action gives the bulk field equations as well as boundary contributions
\be
\d X\(-\pa_\s X + \frac{\pa}{\pa X} \LB\)
\label{el1}
\ee
for the $X$ variation, and
\be
\frac{i}2\(\psi^+\d\psi^+ -\psi^-\d\psi^-\)
\label{el2}
\ee
for the $\psi$ variations.  We assume that there are no boundary
contributions at $t=\pm\infty$. In principle, there could be
$\psi$-dependent terms in the boundary Lagrangian $\LB$, but it turns out that in the absence of a background NS-NS two-form, no such terms arise.

The supersymmetry transformations that leave the action \eqn{act} invariant are\footnote{In our conventions, $\bar\psi=\psi^T(i\g^0)$ and $\g^0=\left(\!\!\begin{array}{cc }0&-1 \\1 &0\end{array}\!\!\right)$,  $\g^1=\left(\!\!\begin{array}{cc}0&1 \\1 & 0 \end{array}\!\!\right)$.}
\be
\begin{array}{ccccc}
\d X &=&\bar\e\psi&\equiv& -i(\e^+\psi^--\e^-\psi^+) \\
\d\psi^+&=&(\g\cdot\pa X \e + F\e)^+&\equiv& -2\e^-\pam X +F\e^+ \\
\d\psi^-&=&(\g\cdot\pa X \e + F\e)^-&\equiv& 2\e^+\pap X +F\e^- \\
\d F &=&\bar\e\g\cdot\pa\psi&\equiv&-2i( \e^+\pap\psi^+ 
+\e^-\pam\psi^-)
\end{array}
\ee
provided we choose the correct boundary Lagrangian and boundary conditions. Under these transformations, the boundary contributions are:
\be
\frac{i}2(\e^+\psi^--\e^-\psi^+)(\pa_\s X -2\frac{\pa\LB}{\pa X})
-\frac{i}2(\e^+\psi^+-\e^-\psi^-)(F+2W')-\frac{i}2(\e^+\psi^-
+\e^-\psi^+)\pa_tX~.
\label{susy}
\ee
We want to find a boundary Lagrangian $\LB$ and boundary conditions on the fields $X,\psi,F$ such that all boundary contributions to both the Euler-Lagrange equations (\ref{el1},\ref{el2}) and the supersymmetry variations (\ref{susy}) vanish. Locality requires that all boundary contributions cancel separately at every boundary; thus it suffices to examine one particular boundary ($\s=0$).

As usual, the $\psi$ boundary contribution to the Euler-Lagrange equation
(\ref{el2}) can vanish only if 
\be
\psi^+(0)=\eta\psi^-(0)~,~~~~\eta=\pm1~.
\label{psibc}
\ee
For (generalized) Neumann boundary conditions\footnote{More precisely, boundary conditions that do not constrain $\pa_tX(0)$.}, the last term in (\ref{susy}) vanishes when
\be
\e^+=-\eta\e^-~.
\label{etabc}
\ee
The remaining boundary contributions from (\ref{el1}) and (\ref{susy}) vanish provided
\be
\pa_\s X(0)=\frac{\pa\LB}{\pa X}(X(0))~,~~
F(0) + 2 W'(X(0)) + \eta\frac{\pa\LB}{\pa X}(X(0)) = 0~.
\label{susybc}
\ee
Supersymmetry invariance of the $\psi$ boundary condition (\ref{psibc}) with respect to the boundary transformation (\ref{etabc}) yields
\be
(\pa_\s X)(0) =\eta F(0)~.
\label{pxfbc}
\ee
The conditions (\ref{susybc},\ref{pxfbc}) determine the boundary Lagrangian $\LB$ in terms of the superpotential $W$
\be
\LB=-\eta W(X(0))
\label{LBW}
\ee
and imply that $F(0)$ satisfies the restriction of the $F$ field equation to the boundary:
\be
F(0)=-W'(X(0))~.
\label{effbc}
\ee
Then the boundary condition for $X$ can be written as
\be
(\pa_\s X)(0)=-\eta W'(X(0))~.
\label{finalxbc}
\ee
Equation \eqn{effbc} is an example of a more general phenomenon: some boundary conditions arise as the restriction of field equations to the boundary. Likewise, supersymmetry invariance of the boundary condition (\ref{effbc}) implies that $\psi$ obeys the restriction to the boundary of the difference of the $\psi^+$ and ($\eta$ times) the $\psi^-$ field equations:
\be
(\pa_\s[\psi^++\eta\psi^-])(0)+\eta W''(X(0))[\psi^++\eta\psi^-](0)=0~,
\label{psifeqbc}
\ee
where we use $(\pa_t[\psi^+-\eta\psi^-])(0)=0$ as a consequence of
(\ref{psibc}). Note that only those field equations that do not determine the time dependence of the boundary fields arise\footnote{The boundary condition (\ref{pxfbc}) is invariant under boundary supersymmetry. The boundary condition (\ref{finalxbc}) gives no new conditions because it is a linear combination of (\ref{pxfbc}) and (\ref{effbc}). Finally we need not check anything for the equation (\ref{LBW}), as it simply defines $\LB$, and is not a boundary condition.}. Indeed, this combination of the field equations is ``conjugate'' to the boundary condition \eqn{psibc} in the sense described in the introduction: in the Euler-Lagrange variation of the action, as we approach the boundary, the boundary condition \eqn{psibc} implies that $\d(\psi^+-\h\psi^-)(0)=0$; since $\d(\psi^+-\h\psi^-)(0)$ multiplies the left-hand side of \eqn{psifeqbc}, when we impose \eqn{psibc}, we must also impose \eqn{psifeqbc} to obtain the complete set of field equations on the boundary. For the bosonic fields $F$ and $X$, the variation of the boundary conditions (\ref{effbc},\ref{finalxbc}) multiply the boundary conditions themselves, and hence these boundary conditions are ``self-conjugate'':
\be
\d_{bose}\Lag=-[\d(\pa_\s X+\eta W')](\pa_\s X+\eta W')+[\d\pa_t X]\pa_t X +[\d(F+W')](F+W')+i(\d W'')\psi^+\psi^-~.
\ee
(The last term vanishes on the boundary because of \eqn{psibc}.)

We emphasize that imposing these constraints on the boundary
does not put us on-shell in the bulk, does not spoil the boundary-value problem for the bulk-field equations, and does not lead to any
peculiarities in defining the path integral. 

Finally, we should vary \eqn{psifeqbc} under supersymmetry, but this gives only time derivatives of \eqn{finalxbc}, and hence no new boundary conditions. Thus we have found the complete orbit of boundary conditions.

All of these results can be extended to the case with an arbitrary
background metric (with the obvious generalization of our notation to more fields $X^i$, etc.) along the lines discussed in \cite{hlz,alz} in the absence of the superpotential, and in \cite{lz}.

The boundary contribution that we have found implies that when the open string propagates through a bulk tachyon field, it feels an extra contribution $\LB=-\eta W(X)$ on the boundary.  We remark that the boundary condition (\ref{finalxbc}) can be interpreted as a BPS condition restricted to the boundary \cite{svv,nsrvn,glvn}; it has also appeared in heat-kernel studies of solitons \cite{bgvnv}.

We next consider Dirichlet boundary conditions; in this case $X(0)= constant$, and hence the $X$ boundary contributions to the Euler-Lagrange equation (\ref{el1}) vanish automatically because $\d X(0)=0$, as does the last term of the boundary contributions to the supersymmetry variations (\ref{susy}). Using \eqn{psibc}, the remaining terms in  (\ref{susy}) vanish if $\e^+=\eta\e^-$ for any $\LB$. This case has been discussed extensively in \cite{alz,Hori,bgvnv} and by many other authors.  These boundary conditions are invariant under supersymmetry transformations, and consequently, further boundary conditions have not been found in the past. However, we find {\em new} boundary conditions that are ``conjugate'' field equations as described above. In contrast with the case of Neumann boundary conditions, these boundary conditions are not implied by any other conditions; nevertheless, they are necessary to ensure consistency of the Euler-Lagrange variational procedure. 

Since the boundary condition \eqn{psibc} on $\psi$ is the same as in the Neumann case, the conjugate field equation remains \eqn{psifeqbc}. However, as the supersymmetry parameter now obeys $\e^+=\eta\e^-$, the supersymmetry variation of \eqn{psifeqbc} is quite different, and leads to the condition
\be
2\e^-\left[\pa^2_\s X +FW'' +\eta\pa_\s(F+W')\right]\!(0)=0~;
\ee
The first two terms are precisely the field equation conjugate to the boundary condition $X(0)= constant$, and lead to the new boundary condition $[\pa^2_\s X +FW''\,](0)=0$; we emphasize that no other considerations lead one to discover this boundary condition.  The remaining terms are the $\s$-derivative of the $F$-field equation, and hence we must impose $[\pa_\s(F+W')](0)=0$ as a new boundary condition. Further supersymmetry variations of these conditions appears to lead to a tower of conditions involving an increasing number of derivatives; as our focus in this section is the Neumann case, we have not worked out the details, but a similar phenomenon is explored with great care in the next section.

We may also find the boundary term $\LB$ from a careful analysis of the superspace action.  We write the action and Lagrangian (\ref{act},\ref{lag}) in $N=(1,1)$ superspace as
\be
I=\frac1\pi \int_\S\frac12(D_+D_--D_-D_+)~ (\frac12D_+\Phi D_-\Phi +iW(\Phi))
\label{superaction}
\ee
where $\Phi$ is a superfield with physical components $\Phi|=X$ and $D_\mp\Phi|=\pm\psi^\pm$ and auxiliary component $F=iD_+D_-\Phi$.  The spinor derivatives obey $\{D_+,D_-\}=0$, $D^2_+=2i\pap$ and $D^2_-=2i\pam$, and we have rewritten the Berezin integral in terms of spinor derivatives. On the boundary, we may introduce a boundary superspace with a single spinor coordinate $\th$ and a single spinor derivative $D=\pa_\th+i\th\pa_t$ obeying $D^2=i\pa_t$; the boundary condition on the supersymmetry parameters (\ref{etabc}) and the usual relation $\d\th^\pm = \e^\pm$ imply that $D_+\Phi(0)= (D +i\th\pa_\s)\Phi(0)$ and $D_-\Phi(0)= -\eta(D -i\th\pa_\s)\Phi(0)$, and hence $(D_+-\eta D_-)\Phi(0)=2D\Phi(0)$.  This is most easily formulated by going to a ``boundary representation''  by introducing a coordinate $\s-i\h\q^+\q^-$ which is annihilated by $D_+-\eta D_-$ (this is analogous to the chiral representation):
\be
\Phi(t,\s,\q^\pm)\equiv\Phi_{bound}(t,\s-i\eta\q^+\q^-,\q^\pm)= \Phi_{bound}(t,\s,\q^\pm)-i\eta\q^+\q^-\pa_\s\Phi_{bound}(t,\s,\q^\pm)
\ee
Substituting this expression for $\Phi$ in \eqn{superaction}, and defining the components in terms of $\Phi_{bound}$, the fermionic integrals give us precisely the component action (\ref{act},\ref{lag}) with the boundary term $\LB=-\eta W(X(0))$. Further, in the presence of a target space Neveu-Schwarz antisymmetric tensor field, we find the correct fermionic boundary terms \cite{hlz,alz2}. 

Boundary terms that arise naturally from superspace measures are familiar in the context $D=4$ super Yang-Mills theory, where the chiral integral $\int d^2\th W^2$ gives rise to a topological term $F\wedge F$; a formal discussion of boundary terms and Berezin integration was first given in \cite{roth}.

\section{BRST invariance of the open bosonic string}
\setcounter{equation}{0}
We now turn to the BRST invariance\footnote{Some aspects of quantization on spaces with boundaries have been considered in \cite{v94,v97}.} of the open bosonic string. The action is 
\be
I=\frac1\pi\int_\S(\Lag_{class} +\Lag_{gh})~,
\label{bose}
\ee
where $\Lag_{class}=2\pap X\pam X$ and the ghost Lagrangian is
\be
\Lag_{gh}=2(\bp\pam\cp+\bm\pap\cmin)~,
\ee
and we do not need an additional boundary term. The $X$ boundary contribution to the Euler-Lagrange equations is simply $-\frac1\pi\d X\pa_\s X$. The ghost contribution is
\be
\frac1\pi(\d\cp\bp -\d\cmin\bm)~.
\ee

The canonical BRST transformations follow from the diffeomorphism invariance of the classical world-sheet action after eliminating the BRST auxiliary field (here $\a=\pbfp, \pbfmm$) 
\beqs
\d X &=& c^\a\pa_\a X\L~,~~~\d c^\a ~=~ -c^\b\pa_\b c^\a \L ~,\nonumber\\
\d \bp&=&\left[(\pap X)^2+2\bp\pap\cp+(\pa_\a\bp)c^\a\right]\L~,
\label{boseBRST}
\eeqs
where $\L$ is a constant anticommuting parameter. The variation of the action (\ref{bose}) under (\ref{boseBRST}) is $\frac1\pi\int\pa_\a\left[c^\a\L (\Lag_{class}+\Lag_{gh})\right]$. When
\be
\pa_\s X(0)=0~,~~~\cp(0)=\cmin(0)~,~~~\bp(0)=\bm(0)~,
\label{bosebc}
\ee
all boundary contributions cancel. However, the BRST transformation of the boundary condition $\pa_\s X =0$ leads to a further condition:
\be
(\pa_\s[\cp+\cmin])(0)=0~.
\label{bosebc2}
\ee
This is the restriction of part of the ghost field equation to the boundary; it is precisely analogous to what we found for the worldsheet spinors $\psi$ in the previous section. Again, only the field equations that do not determine the time dependence of the boundary fields arise. The BRST variations of $\cp(0)-\cmin(0)$ and $\bp(0)-\bm(0)$ vanish.  The BRST variation of (\ref{bosebc2}) vanishes as a consequence of (\ref{bosebc},\ref{bosebc2}) and thus does not give rise to any new conditions.

In the literature, one often finds modified BRST transformations\footnote{These are the BRST transformations for holomorphic {\it operators} $c(z), b(z)$, which are Heisenberg fields, and hence by definition satisfy the field equations. However, as these transformations differ from the canonical ones (\ref{boseBRST}) {\it only} by equation of motion terms that are a separate symmetry of the action, we may use them as alternative transformations of the {\it off-shell} fields $c(\s,t),b(\s,t)$.} that differ from (\ref{boseBRST}) by bulk equation of motion terms ($\pam\cp=\pam\bp=0$):
\be
\tilde\d\cp=-\cp\pap\cp\L~,~~~\tilde\d\bp=\left[(\pap X)^2+ 2\bp\pap\cp+(\pap\bp)\cp\right]\L~.
\ee
Using these transformation laws for the off-shell fields, the boundary contributions to the variation of the action are $-\frac1\pi(\bm\cp\pap\cmin- \bp\cmin\pam\cp)\L$ (using the conditions (\ref{bosebc})).  These vanish only if $(\pa_\s[\cp+\cmin])(0)=0$; as discussed above, this is a restriction of part of the ghost field equation to the boundary. However, the BRST covariance of the boundary condition $\bp(0)=\bm(0)$ imposes a new condition: 
\be
\pa_\s(\bp+\bm)(0)=0~. 
\ee
Varying this condition in turn leads to the conditions
\be
\pa_\s^2(\bp-\bm)(0)=0 ~,~~~\pa_\s^2(\cp-\cmin)(0)=0~.
\ee
Further variations lead to an infinite tower of conditions
\beqs
\pa_\s^{2n}(\bp-\bm)(0)=0 &,&~~~\pa_\s^{2n}(\cp-\cmin)(0)=0~,\nonumber\\
\pa_\s^{2n+1}(\bp+\bm)(0)=0 &,&~~~\pa_\s^{2n+1}(\cp+\cmin)(0)=0~.
\label{bctower}
\eeqs
These conditions can be understood as follows: consider a function $\hat b(\s)$ defined on the double interval $-\pi \le \s \le\pi$ by $\hat b(\s)=\bp(\s)$ for $\s>0$ and $\hat b(\s)=\bm(-\s)$ for $\s<0$, and similarly for $\hat c(\s)$. Then (\ref{bctower}) implies that all the left derivatives of $\hat b(\s)$ at $\s=0$ are equal to all the right derivatives. In other words, $\hat b(\s)$ is smooth at $\s=0$. We can check that the total content of all boundary conditions for the open bosonic string with the improved (holomorphic) BRST transformations is that in the path integral the ghost and antighost fields form closed smooth paths on the double interval.  For the canonical BRST transformations, we did not need this formulation. However, in the next section, we discover the same phenomenon for the spinning string for both sets of BRST transformations.

These problems can be avoided altogether by not eliminating the BRST auxiliary field $d_{\a\b}$. If we keep $d_{\a\b}$, we must also keep the metric $h^{\a\b}$, as $d_{\a\b}$ is the Lagrange multiplier for  the gauge-fixing conditions: $d_{\a\b}(h^{\a\b} -h^{\a\b}_0)$. We also keep the Weyl ghost $c$ as an independent field (see, {\it e.g.,} \cite{pvn}), and find the action
\be
S=\ft1\pi\!\int\!\!\left(-\ft12\sqrt{-h}h^{\a\b}\pa_\a X\pa_\b X + \ft12d_{\a\b}(h^{\a\b}\! - h^{\a\b}_0) +\ft12b_{\a\b}(c^\g\pa_\g h^{\a\b} 
\!-2h^{\a\g}\pa_\g c^\b\! - c\, h^{\a\b})\right),
\ee
as well as the transformations \eqn{boseBRST} for $X$ and $c^\a$, as well as 
\beqs
\d h^{\a\b}\!\!&=&\!\!(c^\g\pa_\g h^{\a\b} 
-h^{\a\g}\pa_\g c^\b - h^{\g\b}\pa_\g c^\a- c\, h^{\a\b})\L~,
\nonumber \\
\d b_{\a\b}\!\!&=&\!\!\L d_{\a\b}~,\,\qquad \d 
d_{\a\b}=0~,\,\qquad \d c = -c^\a\pa_\a c \L ~.
\eeqs
Nilpotency on $b_{\a\b}$ and $d_{\a\b}$ is obvious; on $X$ and $c^\a$, it follows as before; on $c$, it is easy to prove; on $h^{\a\b}$, it is a longer but still straightforward calculation. Under the background diffeomorphism invariance of the action, $b_{\a\b}$ and $d_{\a\b}$ transform as densities (which is why there is no factor of $\sqrt{-h}$ in the ghost and gauge-fixing terms). As usual, the sum of the gauge-fixing and ghost terms is a BRST variation, and hence the nilpotency of the BRST transformations guarantees that there are no boundary terms in the BRST variation of these terms. The classical action is locally Weyl invariant but transforms into a total derivative under Einstein transformations (diffeomorphisms), which leads to a boundary term $\int\pa_\a(c^\a\L\Lag_{class})$ under BRST transformations.

From the field equations for $X^\mu$, the metric $h^{\a\b}$, and the ghost $c^\a$, we find boundary terms proprotional to
\be
(\d X\sqrt{-h} h^{\s\b}\pa_\b X)(0) ~,~~ (\d h^{\a\b} b_{\a\b}c^\s)(0) ~,~~
(b_{\a\b}h^{\s\b}\d c^\a)(0)  ~.
\label{auxterms}
\ee
We can cancel these by choosing the boundary conditions\footnote{Strictly speaking, the last term in \eqn{auxterms} vanishes when $h^{\s\a}b_{\a t}=0$; the last two conditions in \eqn{auxbc} are sufficient to imply this but not clearly necessary. However, the calculation of the orbit of boundary conditions becomes intractable if we only impose $h^{\s\a}b_{\a t}=0$, and there seems to be no inconsistency in choosing the stronger conditions.}
\be
(h^{\s\b}\pa_\b X)(0) = 0 ~,~~c^\s(0)=0 ~,~~ h^{\s t}(0)=0 ~,~~b_{\s t}(0) = 0 ~.
\label{auxbc}
\ee
Note that BRST invariance of the action requires that $(c^\s\Lag_{class})(0)=0$, which is clearly satisfied when $c^\s(0)=0$.  

The first two conditions in (\ref{auxbc}) are BRST invariant; the last condition in (\ref{auxbc}) is the covariantization of $\bp(0) = \bm(0)$ in (\ref{bosebc}), and varying it leads to the new boundary condition $d_{\s t}(0)=0$, which in turn is invariant. Finally, varying $h^{\s t}(0) =0$ lead to the new condition $\pa_\s c^t(0)=0$, which is also invariant. Thus we obtain a finite orbit of boundary conditions. Note that the condition $h^{\s t}(0) =0$ is ``conjugate'' to the condition $d_{\s t}(0)=0$ and $\pa_\s c^t(0)=0$ is ``conjugate'' to $b_{\s t}(0) = 0$ in the sense previously described.

\section{Supergravity in two dimensions}
\setcounter{equation}{0}
The next model whose boundary conditions we study is $N=(1,1)$ supergravity in $1+1$ dimensions. The gauge action is purely topological, and could contribute boundary terms; we do not consider such terms here. 

We focus on the coupling of the supergravity gauge fields $e^a_\a$ and $\chi^\pm_\a$ to the matter fields $X,\psi,$ and $F$. There are four bosonic gauge symmetries--two general coordinate (Einstein) transformations, one local scale (Weyl) transformation, and one local Lorentz transformation--and four fermionic gauge transformations--two supersymmetries and two conformal supersymmetries. These transformations suffice to remove the gauge fields (locally), and hence the number of off-shell bosonic and fermionic degrees of freedom matches; thus, the algebra should close without auxiliary supergravity fields\footnote{For (classically) nonconformal couplings, such as would arise if we coupled the model of section 2 (with a superpotential) to supergravity, a scalar supergravity auxiliary field $S$ is needed.}, which it does as a result of identities that hold only in $1+1$ dimensions\cite{rvn}.

The locally supersymmetric action for the open string is given by
\be
e^{-1}\pi\Lag=-\ft12h^{\a\b}\pa_\a X\pa_\b X-\ft12\bar\psi\g^\a\pa_\a\psi +\ft12 F^2 +(\bar\chi_\a\g^\b\g^\a\psi)\pa_\b X +\ft14(\bar\psi\psi)(\bar\chi_\a\g^\b\g^\a\chi_\b)~.
\label{Lsugra}
\ee
The action $\int\!\Lag$ is invariant (up to boundary terms discussed in detail below) under the following local supersymmetry transformations:
\be
\d X=\bar\e\psi~,~~ \d\psi = \g\cdot\Dl X \e + F\e~,~~ \d F = \bar\e\g\cdot\Dl\psi~,~~ \d e^a_\b=2\bar\e\g^a\chi_\b~,~~ 
\d\chi_\a=D_\a\e~,
\label{Trsugra}
\ee
where $\Dl_\a X\equiv\pa_\a X -\bar\chi_\a\psi$ is the supercovariant derivative of $X$ and $\Dl_\a \psi\equiv D_\a\psi -\g\cdot(\Dl X)\chi_\a-F\chi_\a$ is the supercovariant derivative of $\psi$.  The covariant derivative $D_\a\e$ (and $D_\a\psi$) is given by $D_\a\e\equiv\pa_\a\e+\frac14\o_\a{}^{ab}\g_a\g_b\e$ where $\o_\a{}^{ab}$ is the spin connection with torsion:
\be
\o_\a{}^{ab}\equiv\o_\a{}^{ab}(e)+2(\bar\chi_\a\g^a\chi^b-\bar\chi_\a\g^b\chi^a+\bar\chi^a\g_\a\chi^b)~.
\ee

The action is invariant under local Lorentz, Weyl, and conformal supersymmetry transformations ($\d\chi_\a=\g_\a\e_{CS}$). It is generally (worldsheet) coordinate invariant up to a boundary term $\pa_\s(\xi^\s\Lag)$, which vanishes provided that $\xi^\s=0$ at $\s=0,\pi$.  Local supersymmetry transformations give the following boundary terms:
\be
\d \Lag=\frac1\pi\pa_\s\left[-\frac{e}2\left(\bar\e\g^\s\g^\b\pa_\b X \psi+\bar\psi\g^\s\e  F -(\bar\psi\g^\s\g^\b\e)\ \bar\psi\chi_\b \right) \right] ~.
\label{var1}
\ee
In addition to the boundary terms that arise from local supersymmetry transformations, we have terms that arise from the Euler-Lagrange variations of the action. These vanish provided
\be
[(h^{\s\b}\pa_\b X -\bar\chi_\a\g^\s\g^\a\psi)\d X](0) = 0~,~~ 
[\bar\psi\g^\s\d\psi](0)= 0~.
\label{var3}
\ee
The boundary contribution from the $\psi$ field equation can be rewritten as
\be
\bar\psi\g^\s\d\psi=e^\s_\pbfp\bar\psi\g^\pbfp\d\psi+ e^\s_\pbfm\bar\psi\g^\pbfmm\d\psi=-2i(e^\s_\pbfp\psi^+\d\psi^+ +e^\s_\pbfm\psi^-\d\psi^-)~;
\ee
which vanishes when\footnote{Recall that $e^\s_\pbfpf =\frac12(e^\s_1+e^\s_0), -e^\s_\pbfm =\frac12(e^\s_1-e^\s_0)$, the Minkowski metric is $\eta^{+-}=-2$.}
\be
((e^\s_\pbfp)^{\frac12}\psi^+)(0)= \eta((-e^\s_\pbfm)^{\frac12} \psi^-)(0)~,~~ \eta=\pm1~.
\label{psisugrabc}
\ee
This is the Lorentz-invariant and Weyl-covariant curved worldsheet generalization of $\psi^+(0)=\eta\psi^-(0)$ \eqn{psibc}.  

We now consider the boundary contribution \eqn{var1} from the supersymmetry variation of the action. Since we have found that $\psi^-\propto\psi^+$ on the boundary, the last term vanishes. The first boundary term is proportional to 
\beqs
-\ft12\left(\bar\e\psi h^{\s\b}\pa_\b X\right.  \!\! \!\! &+&  \!\! \!\!  \left.e^\s_ae^\b_b (\bar\e\g^{ab}\psi)\pa_\b X\right)\nonumber \\ &=&
\ft{i}2(\e^+\psi^--\e^-\psi^+) h^{\s\b}\pa_\b X -i(\e^+\psi^-+\e^-\psi^+)(e^\s_{\pbfp}e^\b_{\pbfmm}-e^\s_{\pbfmm}e^\b_{\pbfp}) \pa_\b X~.\nonumber \\ 
\label{var2}
\eeqs
If we require that $\e$ satisfies the curved worldsheet generalization of \eqn{etabc}
\be
((e^\s_\pbfp)^{\frac12}\e^+)(0)=-\eta((-e^\s_\pbfm)^{\frac12}\e^-)(0)~,~~ 
\label{epsugrabc}
\ee
then the second term in \eqn{var2} vanishes. The first term vanishes if
\be
(h^{\s\b}\pa_\b X)(0) -(\bar\chi_\b\g^\s\g^\b\psi)(0)=0~,
\label{hdxbc}
\ee
since terms quadratic in $\psi$ vanish because of \eqn{psisugrabc}.
This is clearly the worldsheet supergravity generalization of the usual Neumann boundary condition, and follows from the Euler-Lagrange variation with respect to $X$ in \eqn{var3}. We are left with the middle term in \eqn{var1}; the boundary conditions on $\psi$ and $\e$ do not cancel  this term, and we are led to conclude that
\be
F(0)=0~,
\label{fsugrabc}
\ee
which is clearly consistent with the $F$ field equation.

We now study the orbits under supersymmetry of the boundary conditions that we have found so far. The supersymmetry variation of \eqn{psisugrabc} has contributions from $\d\psi$ and from $\d e^\s_a$; the boundary condition \eqn{fsugrabc} implies that $\d\psi(0)=\g^a [e_a^\b(\pa_\b X - (\bar\chi_\b\psi))\e](0)$, and hence there are contributions proportional to $\pa_\a X$
\be
[(e^\s_\pbfp)^{\frac12}(-2\e^-e^\a_\pbfm\pa_\a X)](0)- \eta[(-e^\s_\pbfm)^{\frac12}(2\e^+e^\a_\pbfp\pa_\a X)](0)
\ee
as well as contributions proportional to $\chi$, which we discuss below.  Substituting the boundary condition \eqn{epsugrabc} for $\e$, the terms proportional to $\pa_\a X$ become
\be
[(e^\s_\pbfp)(-2\e^+e^\a_\pbfm\pa_\a X)](0)+[(e^\s_\pbfm)(-2\e^+e^\a_\pbfp\pa_\a X)](0)=[\e^+h^{\s\a}\pa_\a X](0)~,
\ee

The variations proportional to $\chi$ come from $\d\psi$
$$
[(e^\s_\pbfp)^{\frac12}(2\e^-(e^\a_\pbfm\bar\chi_\a\psi))](0) - \eta[(-e^\s_\pbfm)^{\frac12}(-2\e^+(e^\a_\pbfp\bar\chi_\a\psi))](0) 
\qquad\qquad\qquad\qquad~~~
$$
\be
\begin{array}{rll}
\qquad\qquad\qquad\qquad\qquad&=& [(e^\s_\pbfp)^{\frac12}(-2i\e^-(e^\a_\pbfm(\chi_\a^+\psi^--\chi_\a^-\psi^+)))](0) 
\\ && - \eta[(-e^\s_\pbfm)^{\frac12}(2i\e^+(e^\a_\pbfp(\chi_\a^+\psi^--\chi_\a^-\psi^+)))](0)~,
\end{array}
\ee
as well as $\d e$
$$
[(e^\s_\pbfp)^{-\frac12}(-\bar\e\g^a\chi_\b e^\s_ae^\b_\pbfp)\psi^+](0) - \eta[(-e^\s_\pbfm)^{-\frac12}(-\bar\e\g^a\chi_\b e^\s_ae^\b_\pbfm)\psi^-](0) \qquad\qquad\qquad\qquad
$$
\be
\begin{array}{rll}
\qquad\qquad\qquad\qquad\qquad &=& [(e^\s_\pbfp)^{-\frac12}(2i)(\e^+\chi^+_\b e^\s_\pbfp+\e^-\chi^-_\b e^\s_\pbfm)e^\b_\pbfp\psi^+](0) \\ 
&& - \eta [(-e^\s_\pbfm)^{-\frac12}(-2i)(\e^+\chi^+_\b e^\s_\pbfp+\e^-\chi^-_\b e^\s_\pbfm)e^\b_\pbfm\psi^-](0)~.
\end{array}
\ee
Combining all these contributions, and substituting the boundary condition \eqn{psisugrabc} on $\psi$ and \eqn{epsugrabc} on $\e$, we obtain \eqn{hdxbc}! Note that all the boundary conditions, (\ref{psisugrabc},\ref{epsugrabc},\ref{hdxbc}) are Lorentz and conformal supersymmetry invariant, as well as Weyl and boundary diffeomorphism covariant.

If one now continues varying the conditions that we have found so far under local supersymmetry, as in the previous section, one finds evidence for an infinite tower of conditions that can be interpreted as arising from an orientifold.  

We note that though the matter boundary conditions get corrected by terms proportional to the supergravity gauge fields, there are {\em no} boundary conditions on the supergravity gauge fields themselves. In retrospect, this is not surprising:  Locally, the supergravity gauge fields can be entirely gauged away.

\section{BRST invariance of the spinning string}
\setcounter{equation}{0}
The ghost structure of the spinning string can be derived in a straightforward way by BRST quantizing {\em all} the local symmetries discussed in the previous section: the 4 bosonic symmetries--Einstein (general coordinate) transformations, local Weyl transformations, and local Lorentz transformations, and the 4 fermionic symmetries--local supersymmetry and local superconformal transformations.  The resulting quantum action depends on 4+4 bosonic and fermionic ghosts and 4+4 bosonic and fermionic antighosts, but one can choose the flat supergravity gauge and integrate out pairs of nonpropagating ghosts and antighosts to obtain the usual flat space Lagrangian:

\beqs
\frac1\pi\Lag_{NSR}&=&\frac2\pi\left(\pap X\pam X +\bp\pam\cp+\bm\pap\cmin\right)+\frac1{2\pi}F^2
\nonumber \\ \nonumber \\
&&+~
\frac{i}\pi\left(\psi^+\pap\psi^++\psi^-\pam\psi^-\right)
-\frac{2i}\pi\left(\bep\pam\c^++\bem\pap\c^- \right)~.
\eeqs

\noindent As in section 2, $\psi^\pm$ are the 2 real components of a worldsheet spinor and $F$ is the real (matter) auxiliary field. Further,  $\c^\pm$ are the imaginary commuting worldsheet spinor supersymmetry ghosts and $\bep,\bem$ are the corresponding real antighosts. As in section 3, the coordinate ghosts $\cp,\cmin$ are real and the coordinate antighosts $\bp,\bm$ are imaginary\footnote{These reality properties differ from those found in some of the literature, and follow from the reality of the Einstein and supersymmetry parameters with the correspondence $\xi=c\L$ and $\e=\c\L$, where $\L$ is the constant anticommuting imaginary BRST parameter. The antighosts transform into real auxiliary fields, $\d b = \L d$ and $\d\b =\L \D$.}. The action is clearly real.

For the matter fields, the BRST transformations are ($\psi_\pm=\mp\psi^\mp$)
\beqs
\d X &=& c^\a\L\pa_\a X + \bar \g \L \psi
=(\cp\pap X +\cmin\pam X  -i(\g^+\psi_++\g^-\psi_-))\L
\nonumber \\ 
\d\psi_\pm & = & c^\a\L\pa_\a\psi_\pm +\frac12(\pa_\bpm c^\bpm\L)\psi_\pm-2(\pa_\bpm X)\g^\pm\L\mp F\g^\mp\L~.
\label{matterbrst}
\eeqs
The terms with $c^\a$ show that $X$ is a scalar and $\psi$ has conformal spin $\frac12$. The transformations proportional to $\g$ are precisely local supersymmetry transformations.

Requiring nilpotency of the BRST laws on $X$ yields the BRST laws for the ghosts:
\beqs
\d\cp &=& c^\a\L\pa_\a\cp-2i\g^+\L\g^+\nonumber\\
\d\cmin &=& c^\a\L\pa_\a\cmin-2i\g^-\L\g^-\nonumber\\
\d\g^+ &=& c^\a\L\pa_\a\g^+-\frac12(\pap\cp\L)\g^+\nonumber\\
\d\g^- &=& c^\a\L\pa_\a\g^--\frac12(\pam\cmin\L)\g^-
\label{ghostbrst}
\eeqs
These results show that $\g$ has conformal spin $-\frac12$ \cite{cft}. 

The BRST rules of $c$ and $\g$ are only nilpotent on-shell; in gauge theories, this typically arises for the antighosts after eliminating the BRST auxiliary fields. Here we have eliminated many nonpropagating ghosts, so it is not surprising that we find nonclosure on the ghosts  and matter fermions as well. Just as for the bosonic string, we can find the ``improved'' BRST transformations of conformal field theory by dropping all nonholomorphic terms and setting $F=0$; these differ from the canonical rules by equation of motion terms, and {\em are} nilpotent off-shell on the ghosts, but, since even the ``improved''  BRST transformations are not holomorphic on $X$, they are not nilpotent on the matter fermions: $\d^2\psi_+\propto 2\g^+\pap(\cmin\pam X-i\g^-\psi_-)$.

After eliminating the BRST auxiliary fields, the Einstein antighosts $b$ transform into the total stress tensor, and the supersymmetry antighosts $\b$ transform into the total supersurrent:
\beqs
\d\bp &=& \L\left[\pap X\cdot\pap X + \frac{i}2\psi_+\pap\psi_++2\bp\pap\cp+(\pa_\a\bp)c^\a\right.\nonumber\\
&&~~~~ -i\left.\left\{\frac32\bep\pap\g^++\frac12(\pap\bep)\g^+\right\}\right]
\nonumber\\\nonumber\\
\d\bep &=& 2\L\left[\psi_+\pap X + 2\bp\g^+-\frac12\left\{(\pa_\a\bep)c^\a+\frac32\bep\pap\cp\right\}\right]~.
\label{antighostbrst}
\eeqs
On-shell, only holomorphic terms remain, again leading to ``improved'' transformations.

We are now ready to apply our analysis of the boundary conditions to the BRST symmetry of the open spinning string. We consider the variation of the action under BRST transformations with a local BRST parameter $\L(\s,t)$, and thus compute the BRST Noether current as well. The variation of the action leads to the following boundary terms:
\beqs
\d S &=&
\int\pam\left[i\psi_+\cmin\L(\pam\psi_+)+2i\psi_+\pap X\g^+\L-2\pap X\pap X \cp\L\right. \nonumber\\
&&~~~~~~~\left.  -2\bp\cmin(\pam\cp)\L-2i\bep\cmin(\pam\g^+)\L\right] \nonumber\\
&&+~\pap\left[ i\psi_+(\pam\psi_+)\cp\L+i\bep\g^+(\pam\cp)\L
-2\bp\cp(\pam\cp)\L\right. \nonumber\\
&&~~~~~~~\left.-2i\bep\cp(\pam\g^+)\L \right]
\nonumber\\
&&+~\pap\L\left[i\cp\psi_+(\pam\psi_+)-2\bp\cp(\pam\cp)+i\bep\g^+(\pam\cp)
\right.\nonumber\\
&&~~~~~~~~~~\left.-2i\bep\cp(\pam\g^+)\right]\nonumber\\
&&+~\pam\L\left[-2\cp\left\{\pap X\pap X+\frac{i}2\psi_+\pap\psi_+ +\bp\pap\cp +\frac12(\pap\bp)\cp\right.\right.\nonumber\\
&&~~~~~~~~~~~~~~~~~~~\left.-i\bep\pap\g^++ \frac{i}4(-(\pap\bep)\g^+ +\bep\pap\g^+)\right\}\nonumber\\
&&~~~~~~~~~~+4i\g^+\left\{\psi_+\pap X+\bp\g^+-\frac14 \bep\pap\cp\right.\nonumber\\
&&~~~~~~~~~~~~~~~~~~~\left.\left.-\frac18(\bep\pap\cp+2(\pap\bep)\cp)\right\} +\frac{i}2\pap(\bep\cp\g^+)
\right]\nonumber\\
&&+~(+\leftrightarrow-)~.
\label{bigbrstvar}
\eeqs
The result shows the following pattern:

\noindent(1) The terms with $\pam\L$ yield the BRST current $j^\pbfm$. We added two terms to complete the ghost stress tensor and two terms to complete the ghost supercurrent; the sum of these extra terms is cancelled by the final total derivative. We thus obtain
\be
j^\pbfm=-2\cp\left[T^{ma}_{\pbfp\pbfp}+\frac12T^{gh}_{\pbfp\pbfp}\right]+ 4i\g^+\left[j^{ma}_{{{\pbfp+}}}+\frac12j^{gh}_{{{\pbfp+}}}\right] +\frac{i}2\pap(\bep\cp\g^+)~,
\ee
where the total derivative is needed to make $j^\pbfm$ a good superconformal tensor. Note that $j^\pbfm$ is holomorphic (it depends only on fields and derivatives with $+$ indices), even though the BRST laws are not holomorphic. 

\noindent(2) The terms with $\pap\L$ appear in exactly the same combination as the the total $\pap$-derivative; they combine to give terms with no derivatives on $\L$\footnote{Although all terms with $\papf\L$ cancel in (\ref{nopa}), the expected $\papf\L\, j^\pbfpf$ contribution arises by considering the terms with $(+\leftrightarrow-)$  in (\ref{bigbrstvar}).}
\beqs
\pap\L[...]+\pap[...]&=&-i\L\pap\left[\psi_+(\pam\psi_+)\cp+\bep\g^+(\pam\cp)
\right.\nonumber\\ &&~~~~~~~~~~~+\left.2i\bp\cp(\pam\cp)- 2\bep\cp(\pam\g^+)\right]~.
\label{nopa}
\eeqs
All these terms are proportional to field equations, and are discussed below.

\noindent(3) The total $\pam$-derivative terms contain both equation of motion terms and terms which do not vanish on-shell. We discuss them separately. Since the boundary is at $\s=0$ (locality requires us to consider the two ends of the string independently), we keep only $\pa_\s$ contributions; we get $\frac12\pa_\s$ from $\pap$ and $-\frac12\pa_\s$ from $\pam$. We also include the terms from the $(+\leftrightarrow-)$ contributions. The terms that do not vanish on-shell are
\be
\pa_\s\left[2\pap X\pap X \cp\L -2i\psi_+\pap X\g^+\L -2\pam X\pam X \cmin\L+ 2i\psi_-\pam X\g^-\L\right]~.
\ee
Using the boundary conditions $\cp(0)=\cmin(0),~\pa_\s X(0)=0$, we see that we must impose $\psi_+(0)\g^+(0)=\psi_-(0)\g^-(0)$. This is satisfied by 
\be
\psi_+(0)=-\h\psi_-(0)~,~~\g^+(0)=-\h\g^-(0)~,~~\h=\pm1~;
\ee
these boundary conditions are compatible with the conditions that we found for rigid supersymmetry in section 2.

The boundary terms that are proportional to field equations all cancel when we impose $\cp(0)=\cmin(0)$ except for
\be
\pap(i\bep\g^+\pam\cp\L)+(+\leftrightarrow-)~.
\ee
This term vanishes provided $\bep\g^+(0)=\bem\g^-(0)$ and $\pa_\s(\cp+\cmin)(0)=0$.  We impose these boundary conditions; we recover them below when we consider the closure of the boundary conditions with respect to BRST variations.

The extra boundary conditions that arise from the ghost Euler-Lagrange variation of the action are
\be
\bp(0)=\bm(0)~,~~\bep(0)=-\eta\bem(0)~.
\ee

We now study the further conditions needed to ensure that the boundary conditions that arose from imposing BRST invariance of the action and consistency of the Euler-Lagrange variations are themselves invariant under BRST transformations.  There are seven conditions that we have imposed so far:
$$
\pa_\s X(0)=0~;~~ (\cp-\cmin)(0)=0 ~;~~ (\bp-\bm)(0)=0~;~~ \pa_\s(\cp+\cmin)(0)=0~;
$$
\be
(\psi_++\eta\psi_-)(0)=0~;~~ (\g^++\eta\g^-)(0)=0~;~~(\bep+\eta\bem)(0)=0~.
\ee
Except for $\pa_\s(\cp+\cmin)(0)=0$, these relations are plausible and could have been guessed at the outset. They are preserved provided the following new boundary conditions hold:
\be
F(0)=0~;~~ \pa_\s(\psi_+-\eta\psi_-)(0)=0 ~;~~ 
 \pa_\s(\g^+-\eta\g^-)(0)=0 ~;~~ \pa_\s(\bep-\eta\bem)(0)=0~,
\label{varlevel1}
\ee
where $F$ is the matter auxiliary field. At this level, we do {\em not} find the boundary condition $\pa_\s(\bp+\bm)(0)=0$ because it is multiplied by $(\cp-\cmin)(0)$, which vanishes.  The condition $F(0)=0$ was obtained in section 2 by requiring invariance of the classical action under supersymmetry in the absence of a superpotential; now it is obtained from BRST invariance, but, of course, the two are closely related. (Usually, one eliminates $F$, but we have kept it for completeness.)

Further variation of (\ref{varlevel1}) lead to two new conditions:
\be
\pa^2_\s(\cp-\cmin)(0)=0~;~~\pa_\s(\bp+\bm)(0)=0~;
\ee
note the expected appearance of the condition $\pa_\s(\bp+\bm)(0)=0$.

The pattern is clear: the canonical BRST transformation rules of the spinning string require the same infinite tower of boundary conditions as the improved BRST transformations for the bosonic string! As one goes up the tower, the conditions on the bosonic and fermionic sectors alternate; at each level, the number of $\pa_\s$'s increases and a relative sign changes. For example, the variation of $\pa_\s(\bp+\bm)(0)=0$ gives the new conditions\footnote{The variation of $\pa^2_\s( c^\pbfpf-c^\pbfmf)(0) \equiv2\pa_\s^2c^\s(0)=0$ does {\em not} give a new condition because $\d c^\s= c^\s\pa_\s c^\s +c^t\pa_t c^\s$ and $c^\s(0)=\pa_\s c^t(0) =\pa_\s^2 c^\s(0)=0$.}
\be
\pa^2_\s(\psi_++\eta\psi_-)(0)=0~;~~  \pa^2_\s (\g^++\eta\g^-)(0)=0~;~~
\pa^2_\s (\bep+\eta\bem)(0)=0~.
\ee
Variation of $\pa^2_\s(\psi_++\eta\psi_-)(0)=0$ yields $\pa^2 F(0)=0$ and 
$\pa^3_\s X(0)=0$, and obvious further conditions; no condition $\pa_\s^2 X(0)=0$ or $\pa_\s F(0)=0$ arises. 

The complete set of boundary conditions is:
$$
\pa^{2n+1}_\s X(0)=0~;~~ \pa^{2n}_\s F(0)=0~;~~ 
\pa^{n}_\s (\cp -(-)^n\cmin)(0)=0~;~~
\pa^{n}_\s (\bp -(-)^n\bm)(0)=0~;~~
$$
\be
\pa^{n}_\s (\psi_+ +(-)^n\eta\psi_-)(0)=0~;~~ 
\pa^{n}_\s (\g^+ +(-)^n\eta\g^-)(0)=0~;~~ 
\pa^{n}_\s (\bep +(-)^n\eta\bem)(0)=0~.
\label{finalbc}
\ee
These conditions hold for for $n=0,1,2,\dots$ and for all world-sheet time $t$.

We now interpret our results geometrically. We define fields on the double interval $-\pi\le\s\le\pi$ by:
$$
\hat X(\s,t)=\left\{
\begin{array}{ll}
X(\s,t) & \s\ge 0  \\ 
X(-\s,t) & \s\le 0
\end{array}
\right.
~;~~~~
\hat F(\s,t)=\left\{
\begin{array}{ll}
~~F(\s,t) & \s\ge 0  \\ 
-F(-\s,t) & \s\le 0
\end{array}
\right. ~;
$$
$$
~\hat c(\s,t)=\left\{
\begin{array}{ll}
\cp(\s,t) & \s\ge 0  \\ 
\cmin(-\s,t) & \s\le 0
\end{array}
\right.
~;~~~~~
\hat b(\s,t)=\left\{
\begin{array}{ll}
\bp(\s,t) & \s\ge 0  \\ 
\bm(-\s,t) & \s\le 0
\end{array}
\right. ~;
$$
$$
~~~\,\hat \psi(\s,t)=\left\{
\begin{array}{ll}
~\psi_+(\s,t) & \s\ge 0  \\ 
-\eta\psi_-(-\s,t) & \s\le 0
\end{array}
\right.\!
;~
\hat \g(\s,t)=\left\{
\begin{array}{ll}
~\g^+(\s,t) & \s\ge 0  \\ 
-\eta\g^-(-\s,t) & \s\le 0
\end{array}
\right.~;
$$
\be
~~~\,\hat \b(\s,t)=\left\{
\begin{array}{ll}
~\bep(\s,t) & \s\ge 0  \\ 
-\eta\bem(-\s,t) & \s\le 0
\end{array}
\right.
~.\qquad\qquad\qquad\qquad\qquad\qquad~
\ee
\vskip 10pt
\noindent The conditions (\ref{finalbc}) can all be interpreted as smoothness conditions on the fields defined on the double interval, that is, left-derivatives are equal to right-derivatives at the boundary. Specifically, they imply $\hat X$ is a smooth {\em symmetric} function, $\hat F$ is a smooth {\em antisymmetric} function, and $\hat b, \hat c, \hat\psi, \hat \g, \hat\b$ are all smooth at the boundaries without any particular symmetry properties in the bulk. Thus in a path integral approach, BRST-invariance leads us to consider smooth fields on the double interval. (Of course, this is the usual dense set of paths that physicists always consider; strictly speaking, most paths are not even differentiable, but since the BRST transformations as well as the Lagrangian involve derivatives, that would require an entirely different approach.)

We can again consider the ``improved'' BRST transformation rules found by dropping nonholomorphic terms in (\ref{matterbrst},\ref{ghostbrst},\ref{antighostbrst}). These holomorphic BRST transformations have the same infinite set of boundary conditions with the same geometrical interpretation as the canonical BRST transformations.

With hindsight, it is merely an accident (or a miracle?) that the canonical BRST transformation rules for the bosonic string did not lead to an infinite set of boundary conditions; indeed, the higher-order boundary conditions on the bosonic sector of the spinning string follow from imposing BRST-invariance of the boundary conditions for the fields in the {\em fermionic} sector.

A certain exhaustion prevents us from investigating the effect of BRST auxiliary fields in the spinning (NSR) string.

\section{Symmetries of the superstring}
\setcounter{equation}{0}
As a last example, we consider the Green-Schwarz superstring and study the compatibility of the boundary conditions with $\k$(Siegel)-symmetry. We remind the reader that the Green-Schwarz superstring is an alternative formulation of string theory in terms of worldsheet fields which are spacetime superspace coordinates $X^\mu(t,\s)$ and $\q^i(t,\s)$ ($i=1,2$); this has the advantage that it makes spacetime supersymmetry manifest, though it is difficult to quantize covariantly. There are two pure closed superstring theories: the IIA theory, in which the Weyl spinor coordinates $\q^i(t,\s)$ have opposite chirality, and the IIB theory, in which they have the same chirality. There is in addition the type I theory, which has both a closed and an open string sector, and has two spacetime Weyl spinors $\q^i(t,\s)$ with the same chirality. In all of these theories, there appear to be too many fermionic fields; as was discovered by Siegel for the superparticle \cite{ws1} and the superstring in D=3 \cite{ws2}, and extended to D=10 by Green and Schwarz \cite{gs}, half of these can be gauged away by a local fermionic gauge symmetry called $\k$-symmetry.  Here we focus on $\k$-invariance of the open sector of the type I string. The $\k$-transformations  of $X^\mu(t,\s)$ and $\q^i(t,\s)$ are
\be
\d X^\mu=-\sum_{i=1}^2\d\bar\q^i\G^\mu\q^i~, ~~~\d\q^i=\pi^\mu_\a\G_\mu\k^{i\a}~,
\ee
where the parameters anticommuting $\k^i$ are (anti)self-dual:
$$
\k^{1\a}=P_-^{\a\b}\k^{1}_\b~, ~~\k^{2\a}=P_+^{\a\b}\k^{2}_\b~, ~~\bar\k^{1\a}=\bar\k^{1}_\b P^{\b\a}_+~,~~
\bar\k^{2\a}=\bar\k^{2}_\b P^{\b\a}_-~;
$$
\be
~P^{\a\b}_\pm=\frac12(h^{\a\b}\pm\frac{\e^{\a\b}}{\sqrt{-h}})=P^{\b\a}_\mp~,
\label{kappadual}
\ee
and
\be
\pi^\mu_\a=\pa_\a X^\mu -\sum_{i=1}^2\bar\q^i\G^\mu\pa_\a\q^i~.
\ee
In the projector, $h$ is the determinant of the worldsheet metric $h_{\a\b}$;
the combination $\sqrt{-h}h^{\a\b}$, which is all we need, transforms as\footnote{Though we work in conformal gauge $\sqrt{-h}h^{\a\b}=\eta^{\a\b}$ below, for now we keep the metric $h$ arbitrary.}
\be
\d(\sqrt{-h}h^{\a\b})=-8\sqrt{-h}(\bar\k^{1\a}P_-^{\b\c}\pa_\c\q^1 +\bar\k^{2\a}P_+^{\b\c}\pa_\c\q^2)~.
\label{hkappa}
\ee
Though it is not obvious, because of the (anti)self-duality of $\k^i$ and the relation $\e_{\a\b}P^{\b\g}_\pm=\pm\sqrt{-h}h_{\a\b}P^{\b\g}_\pm$, the right-hand side of (\ref{hkappa}) is symmetric and traceless. The (anti)self-duality relations (\ref{kappadual}) can be rewritten as
\be
\k^1_t = \sqrt{-h}\k^{1\s}~,~~\k^2_t = -\sqrt{-h}\k^{2\s}~,~~\k^1_\s= -\sqrt{-h}\k^{1t}~,~~\k^2_\s= \sqrt{-h}\k^{2t}~.
\label{kappaexplicit}
\ee

The worldsheet action of the Green-Schwarz superstring is\footnote{If one writes the second term with $\pa_\a X^\mu$ instead of $\pi_\a^\mu$, the sign of the last term changes.}
\be
S=\frac{-1}\pi\int \frac12\sqrt{-h}h^{\a\b}\pi^\mu_\a\pi^\nu_\b\eta_{\mu\nu} + 
\e^{\a\b}\left[ \pi^\mu_\a(\bar\q^1\G_\mu\pa_\b\q^1-\bar\q^2\G_\mu\pa_\b\q^2) - (\bar\q^1\G^\mu\pa_\a\q^1)(\bar\q^2\G_\mu\pa_\b\q^2) \right] .
\ee
This action is invariant under local Weyl rescalings $\d h_{\a\b} = \l h_{\a\b}, ~\d X=\d\q^i=0$, rigid spacetime supersymmetry transformations $\d \q^i = \e^i, ~\d X^\mu=\sum_{i=1}^2\d\bar\q^i\G^\mu\q^i, ~\d h_{\a\b}=0$, as well as $\k$-transformations and worldsheet diffeomorphisms.  The diffeomorphisms can be used to go to conformal gauge $\sqrt{-h}h^{\a\b}=\eta^{\a\b}$, which leads to nonlocal compensating transformations in the $\k$-transformations that we discuss below.

From the Euler-Lagrange equations, one finds the following boundary conditions:
\be
(\q^1-\q^2)(0)~,~~(h^{\s\a}\pi^\mu_\a)(0)\equiv\pi^{\s\mu}(0)=0~.
\label{kbc}
\ee 
(In the light-cone gauge, $\pi^\mu_\s=\pa_\s X^\mu$ for the transverse coordinates, and hence we recover the usual Neumann boundary conditions.)
Supersymmetry transformations of the action lead to a boundary term that vanishes when $\e^1=\e^2$, consistent with the boundary condition $\q^1(0)=\q^2(0)$.

We now consider the orbit of successive symmetry variations of the boundary conditions \eqn{kbc}. Under diffeomorphisms, we find
\be
\d(\q^1-\q^2)(0)=(\xi^\a\pa_\a(\q^1-\q^2))(0)~,~~\d\pi^{\s\mu}(0)= (\xi^\a\pa_\a\pi^{\s\mu}-\pi^{\a\mu}\pa_\a\xi^\s)(0)~.
\ee
Though in principle one could consider other conditions, the only consistent boundary condion that makes these vanish when we impose \eqn{kbc} is ({\it cf.} the comment above \eqn{var1})
\be
\xi^\s(0)=0~.
\label{xis}
\ee

A direct construction of  the orbit of successive $\k$-symmetry variations of the boundary conditions \eqn{kbc} in arbitrary diffeomorphism gauge rapidly becomes very complicated. We therefore restrict ourselves to conformal gauge. The $\k$-transformations of the metric \eqn{hkappa} do not preserve the conformal gauge condition, and hence must be augmented by compensating diffeomorphisms that restore the gauge. If the compensating diffeomorphisms can be chosen to obey \eqn{xis}, then they will preserve the boundary conditions \eqn{kbc}, and will not be needed in the $\k$-transformations of the boundary conditions. This observation leads to great simplifications in subsequent calculations. Before explicitly constructing the compensating transformations, we study the $\k$-transformations of the boundary conditions \eqn{kbc} simply assuming \eqn{xis} and ignoring the compensating transformations.

Under a $\k$-transformation, 
\be
\d(\q^1-\q^2) = \pi^\mu_\a\G_\mu(\k^{1\a}-\k^{2\a}) \equiv\pis_\a(\k^{1\a}-\k^{2\a})~.
\ee
Using the boundary condition $\pi^{\s\mu}(0)=0$, this implies
\be
(\k^1_t-\k^2_t)(0)=0~;
\label{kappabc}
\ee
the (anti)self-duality of $\k^i$ ({\it cf.} \ref{kappaexplicit}) then implies
\be
(\k^{1\s}+\k^{2\s})(0)=0~.
\ee
To study the variation of the boundary condition $\pi^{\s\mu}(0)=0$, 
we use 
\be
\d\pi_\a^\mu=2\pa_\a\bar\q^i\G^\mu\pis_\b\k^{i\b}~.
\ee
In conformal gauge, we find
\be
\d\pi^\mu_\s(0) =2\left(\pa_\s(\q^1+\q^2)\G^\mu\pis_t\k^{1t}\right)\!(0)~.
\label{pivar}
\ee
This implies the boundary condition
\be
\pa_\s(\q^1+\q^2)(0)=0~.
\label{dthetabc}
\ee
We now show that (\ref{dthetabc}) follows from a combination of the $\q$ field equations restricted to the boundary.

The $\q$ field equations are
\beqs
\frac\d{\d\bar\q^1}S &=& (\G^\mu\q^1)\frac\d{\d X^\mu}S+\frac4\pi\sqrt{-h}P^{\a\b}_-\pis_\a\pa_\b\q^1~,\nonumber\\
\frac\d{\d\bar\q^2}S &=& (\G^\mu\q^2)\frac\d{\d X^\mu}S+\frac4\pi\sqrt{-h}P^{\a\b}_+\pis_\a\pa_\b\q^2~,
\eeqs
where
\be
\frac\d{\d X^\mu}S=\pa_\a\left[ \pi^{\a\mu}+\e^{\a\b}(\bar\q^1\G^\mu\pa_\b\q^1- \bar\q^2\G^\mu\pa_\b\q^2)\right]~.
\ee
We consider $\left(\frac\d{\d\bar\q^1}S - \frac\d{\d\bar\q^2}S\right)\!(0)$.
The terms proportional to $\frac\d{\d X^\mu}S$ cancel because $(\q^1 -\q^2)(0)=0$. The rest is proportional to 
\be
\left(\sqrt{-h}\pis^\b\pa_\b(\q^1-\q^2)-\e^{\a\b}\pis_\a\pa_\b(\q^1+\q^2)\right)\!(0)~.
\ee
The first term vanishes because $\pi^\s(0)=0$ and $(\q^1-\q^2)(0)=0$.
The remaining term is $-\left(\pis_t \pa_\s(\q^1+\q^2)\right)\!(0)$, which is proportional to the boundary condition (\ref{dthetabc}), and thus (\ref{dthetabc}) is proportional to the combination of field equations $\left(\frac\d{\d\bar\q^1}S - \frac\d{\d\bar\q^2}S\right)\!(0)$.

The $\k$-invariance of the boundary condition $\pi^{\s\mu}(0)=0$ has forced us to add a new condition (compatible with the equations of motion): $(\pa_\s(\q^1+\q^2))(0)=0$. We now vary this condition and see if these conditions close. The variation gives
\be
\pa_\s(\pis_\a(\k^{1\a}+\k^{2\a}))(0)= 
((\pa_\s\pis_t)(\k^{1t}+ \k^{2t}))(0)+
(\pis_t\pa_\s(\k^{1t}+\k^{2t}))(0)~,
\label{dthetavar}
\ee
where we have used $(k^{1\s}+\k^{2\s})(0)=0$ to simplify the first term and $\pi^\mu_\s(0)=0$ to simplify the second term. The first term vanishes as a consequence of previous boundary conditions:
\beqs
(\pa_\s\pis_t)(0)&=&(\pa_t\pis_\s)(0)+2\G_\mu(\pa_t\bar\q^i \G^\mu\pa_\s\q^i)(0)\nonumber\\
&=&(\pa_t\pis_\s)(0)+2\G_\mu(\pa_t\bar\q^1 \G^\mu\pa_\s(\q^1+\q^2))(0)~,
\label{paspist}
\eeqs
which vanishes as a consequence of $\pi^\mu_\s(0)=0$ and $(\pa_\s(\q^1+\q^2))(0)=0$ (we used $(\q^1-\q^2)(0)=0$ to simplify the last term in second line of \eqn{paspist}).
Finally, the last term in \eqn{dthetavar} vanishes when we impose a new condition on the $\k$-parameter:
\be
(\pa_\s(\k^1_t+\k^2_t))(0)=0.
\label{dkappabc}
\ee
With this condition, the boundary conditions are preserved by both diffeomorphisms and $\k$-transformations in conformal gauge {\em provided} that the compensating diffeomorphism satisfies $\xi^\s(0)=0$.

We now study the compensating diffeomorphism.
The boundary conditions that are imposed on the diffeomorphism parameters by the conformal gauge choice follow from
\be
\d\sqrt{-h}h^{\a\b}(0)=\left(\xi^\c\pa_\c(\sqrt{-h}h^{\a\b}) - \sqrt{-h}h^{\c(\a}\pa_\c\xi^{\b)} +(\pa_\c\xi^\c)\sqrt{-h}h^{\a\b}\right)\!(0)~.
\label{hdiff}
\ee
These vanish in conformal gauge when $\xi^\s(0)=0$ is satisfied as well as
\be
\pa_\s\xi^t(0)=0 ~,~~ (\pa_t\xi^t-\pa_\s\xi^\s)(0)=0~.
\label{xit}
\ee

In the remainder of this paper, we find the compensating transformation and demonstrate that it can be chosen to obey $\xi^\s(0)=0$.  These calculations are most readily performed in the coordinate basis $\pap\equiv\half(\pa_t+\pa_\s), ~\pam\equiv\half(\pa_t-\pa_\s)$. In this basis, we have
\be
\eta^{\pbfp\pbfmm}=-2~,~~P_+^{\pbfp\pbfmm}=P_-^{\pbfmm\pbfp}= -2~,
\ee
which implies that the duality conditions (\ref{kappadual},\ref{kappaexplicit}) on $\k$ become simply
\be
\k^{1\pbfp}=\k^{2\pbfmm}=0~.
\ee
The conditions that the compensating diffeomorphism $\xi_{comp}$ satisfies follow from \eqn{hkappa} and \eqn{hdiff}:
\be
8\bar\k^{1\pbfmm}\pap\q^1+\pap\xi_{comp}^\pbfmm=0~,~~
8\bar\k^{2\pbfp}\pam\q^2+\pam\xi_{comp}^\pbfp=0~.
\ee
These integrate to
\beqs
\xi_{comp}^\pbfmm(t,\s)&=& -8\int^{t+\s}\! ds\left(\bar\k^{1\pbfmm}\pap\q^1\right)\!\left(\half(s+t-\s),\half(s-t+\s)\right)+\xi_0^\pbfmm(t-\s)~,\nonumber\\\nonumber\\
\xi_{comp}^\pbfp(t,\s)&=& -8\int^{t-\s}\! ds\left(\bar\k^{2\pbfp}\pam\q^2\right)\!\left(\half(s+t+\s),\half(-s+t+\s)\right)+\xi_0^\pbfp(t+\s)~,\nonumber\\
\eeqs
where $\xi_0$ are residual diffeomorphisms that preserve conformal gauge (conformal transformations). 

We now consider the condition \eqn{xis}, which in this coordinate basis becomes\footnote{Until now, we have used the notation that for any field or parameter $\Psi(t,\s)$, $\Psi(0)$ is the value at the boundary (either $\s=0$ or $\s=\pi$); we now abandon this convention, and indicate the arguments explicitly.}
\be
\xi^\pbfp(t,0)=\xi^\pbfmm(t,0)~,~~\xi^\pbfp(t,\pi)=\xi^\pbfmm(t,\pi)~;
\ee
since the compensating diffeomorphism is nonlocal, here we have to consider both the boundaries at $\s=0$ and at $\s=\pi$ explicitly. At $\s=0$, 
\beqs
&&-8\int^{t}\! ds\left(\bar\k^{1\pbfmm}\pap\q^1\right)\!\left(\half(s+t),\half(s-t)\right) +\xi_0^\pbfmm(t) \qquad\qquad\qquad\qquad\qquad \nonumber\\\nonumber\\
&&\qquad\qquad\qquad = -8\int^{t}\! ds\left(\bar\k^{2\pbfp}\pam\q^2\right)\!\left(\half(s+t),\half(-s+t)\right)+\xi_0^\pbfp(t)~.
\label{xibc0}
\eeqs
We satisfy this condition by choosing $\xi_0^\pbfp(t)-\xi_0^\pbfmm(t)$.
We are left with the boundary condition at $\s=\pi$
$$
-8\int^{t+\pi}\! ds\left(\bar\k^{1\pbfmm}\pap\q^1\right)\!\left(\half(s+t-\pi),\half(s-t+\pi)\right) +\xi_0^\pbfmm(t-\pi) 
\qquad\qquad\qquad\qquad
$$
\be
\qquad\qquad\quad = -8\int^{t-\pi}\! ds\left(\bar\k^{2\pbfp}\pam\q^2\right)\!\left(\half(s+t+\pi),\half(-s+t+\pi)\right)+\xi_0^\pbfp(t+\pi)~.
\label{xibcpi}
\ee
Using \eqn{xibc0} at a time $t+\pi$, this may be rewritten as
\be
8\int^{t+\pi}_{t-\pi}\! ds\left(\bar\k^{2\pbfp}\pam\q^2\right)\!\left(\half(s+t+\pi),\half(-s+t+\pi)\right)+\xi_0^\pbfp(t-\pi)-\xi_0^\pbfp(t+\pi)=0~.
\label{xibcpi2}
\ee
The integral is just some function of $t$, and we can always find a
$\xi_0^\pbfp(t)$ that satisfies this finite difference equation. (One way to see this is to Taylor expand the difference; the matrix relating the Taylor coefficients of the difference to the Taylor coefficients of $\xi_0^\pbfp(t)$ is invertible.)\  This completes our demonstration that the compensating diffeomorphism satisfies the boundary condition $\xi_{comp}^\s(0)=0$.

\vspace{5mm}
\noindent{\bf{Acknowledgement}}
\vspace{3mm}

\noindent{The work of UL is supported in part by VR grant 5102-20005711 and by EU contract HPRN-C7-2000-0122. The work of MR and PvN is supported in part by NSF Grant No. PHY-0098527.}
\newpage

\end{document}